\documentclass[twocolumn]{aastex631}
\usepackage{nicefrac}
\usepackage{amssymb}
\usepackage{latexsym}
\usepackage{amsmath, mathrsfs, bm}
\usepackage{eqnarray}
\usepackage{url}
\usepackage{cases}
\usepackage{epsfig}
\usepackage{color}
\usepackage{graphicx}
\usepackage{grffile}
\usepackage{float}
\usepackage{soul}
\usepackage{enumerate}
\usepackage{multirow}
\usepackage{threeparttable}
\usepackage{xspace}

\usepackage{bm}
\usepackage{times}

\usepackage{etoolbox}
\usepackage{multirow}

\begin{document}


\title{Model-independent $H_0$ within FLRW: Joint constraints from GWTC-3 standard sirens and strong lensing time delays}

\author[0009-0003-8111-0470]{Ji-Yu Song}
\affiliation{Key Laboratory of Cosmology and Astrophysics (Liaoning Province) \& Department of Physics, College of Sciences, Northeastern University, Shenyang 110819, China}

\author[0000-0001-7492-874X]{Jing-Zhao Qi}
\affiliation{Key Laboratory of Cosmology and Astrophysics (Liaoning Province) \& Department of Physics, College of Sciences, Northeastern University, Shenyang 110819, China}

\author[0000-0002-3512-2804]{Jing-Fei Zhang}
\affiliation{Key Laboratory of Cosmology and Astrophysics (Liaoning Province) \& Department of Physics, College of Sciences, Northeastern University, Shenyang 110819, China}

\author[0000-0002-6029-1933]{Xin Zhang}
\affiliation{Key Laboratory of Cosmology and Astrophysics (Liaoning Province) \& Department of Physics, College of Sciences, Northeastern University, Shenyang 110819, China}
\affiliation{Key Laboratory of Data Analytics and Optimization for Smart Industry (Ministry of Education), Northeastern University, Shenyang 110819, China}
\affiliation{National Frontiers Science Center for Industrial Intelligence and Systems Optimization, Northeastern University, Shenyang 110819, China}

\correspondingauthor{Jing-Zhao Qi}
\email{qijingzhao@mail.neu.edu.cn}
\correspondingauthor{Xin Zhang}
\email{zhangxin@mail.neu.edu.cn}

\begin{abstract}

We use 47 gravitational-wave (GW) standard sirens from the third Gravitational-Wave Transient Catalog to calibrate distances in the strong gravitational lensing (SGL) system RXJ1131-1231 and constrain the Hubble constant ($H_0$) via the distance sum rule, without assuming a specific cosmological model. For $\Omega_K = 0$, we obtain $H_0 = 73.22^{+5.95}_{-5.43}$ ${\rm km}~{\rm s}^{-1}~{\rm Mpc}^{-1}$ and $H_0 = 70.40^{+8.03}_{-5.60}$ ${\rm km}~{\rm s}^{-1}~{\rm Mpc}^{-1}$ by breaking the mass-sheet transform using lens galaxy's mass models and stellar kinematics, respectively. Allowing $\Omega_K$ to vary increases the central value of $H_0$ and reduces its precision. We find that GW dark sirens have significant potential for calibrating SGL systems, due to their relatively higher redshifts. By combining 42 binary black holes and RXJ1131-1231, we obtain an $H_0$ constraint with a precision approximately 40\% better than the measurement from GW170817 using the Hubble law. This suggests that high-precision, model-independent $H_0$ measurements can be achieved with this method as the redshift range of GW dark sirens expands, even without the need for GW bright sirens.

\end{abstract}

\keywords{Cosmology (343) --- Hubble constant (758) --- Gravitational waves (678) --- Strong gravitational lensing (1643)}

\section{Introduction}

The $\Lambda$CDM model is successful, as it explains the most of cosmic observations and fits the cosmic microwave background (CMB) data with stunning precisions. However, the tensions in some parameter measurements from different observations challenge the internal consistency of the $\Lambda$CDM model, especially the Hubble tension. The Hubble constant ($H_0$) inferred by globally fitting the $\Lambda$CDM model to the Planck CMB measurement is in over 5$\sigma$ tension with that measured with the SH0ES (Supernovae and $H_0$ for the Equation of State of dark energy) distance ladder \citep{Planck:2018vyg, Riess:2021jrx}. Determination of the cause of the Hubble tension is significant, as it could hint at the new physics beyond the $\Lambda$CDM model. Currently, no extended cosmological model can resolve the Hubble tension and agree well with observations at the same time \citep{Guo:2018ans,DAmico:2020ods,Hill:2020osr}, while numerous studies demonstrated that systematics in observations may not explain the serious discrepancy of $H_0$ \citep{Planck:2016tof,Jones:2018vbn,Kenworthy:2019qwq,Riess:2019cxk,Riess:2021jrx,Castello:2021uad,Planck:2019evm}, including using the high-precision James Webb Space Telescope \citep{Riess:2024vfa,Li:2024pjo,Pascale:2024qjr,Li:2024yoe,Riess:2024ohe}. However, $H_0$ values from distance ladders using different calibrators and distance indicators are inconsistent \citep{Freedman:2024eph,Lee:2024qzr}, pushing continuous efforts to search for potential systematical bias in SH0ES measurements \citep{Huang:2024erq,Huang:2024gfw,Perivolaropoulos:2024yxv,Hogas:2024qlt}. A cosmological-model-independent measurement of $H_0$ from the third-party observations undoubtedly helps resolve the Hubble tension.

Recently, based on the distance sum rule in the Friedmann-Lema\^{i}tre-Robertson-Walker (FLRW) metric \citep{Rasanen:2014mca}, \citet{Collett:2019hrr} used Type Ia supernovae (SNe Ia) and strong gravitational lensing time delays (SGLTDs) to measure $H_0$ and $\Omega_K$ without assuming a specific cosmological model. In a strong gravitational lensing (SGL) system, the time delay is related to the time-delay distance, which is a ratio of three angular distances: the observer to the lens galaxy, the observer to the source, and the lens galaxy to the source. In the FLRW metric, these three distances are interconnected through the distance sum rule, incorporating $H_0$ and $\Omega_K$. By using SNe Ia as standard candles to calibrate these distances, one can constrain $H_0$ and $\Omega_K$ without reliance on specific cosmological models. However, unanchored SNe Ia can only measure relative distances, and measuring absolute distances with SNe Ia requires distance ladders to calibrate their absolute magnitude, $M_B$. Consequently, when using unanchored SNe Ia to determine $H_0$ with this method, $H_0$ becomes fully degenerate with $M_B$, potentially introducing significant uncertainties. Subsequently, \citet{Wei:2020suh} used the ultraviolet versus X-ray luminosity correlation of quasars to calibrate distances in SGL systems and measure $H_0$ and $\Omega_K$. \citet{Qi:2022sxm} made a forecast using strongly lensed SNe Ia within this method.

Gravitational-wave (GW) standard sirens provide self-calibrated luminosity distance measurements, serving as more reliable distance indicators. GWs from compact binary coalescences directly encode luminosity distances in their waveform amplitudes. By obtaining the redshifts of GW sources, it is possible to constrain the history of cosmic expansion \citep{Schutz:1986gp}. Detecting the electromagnetic (EM) counterparts of GW sources can help identify their host galaxies and determine the redshifts \citep{Schutz:1986gp,LIGOScientific:2017adf,LIGOScientific:2017vwq,LIGOScientific:2017zic}. Depending on the detection of EM counterparts, GW standard sirens are classified as ``bright sirens'' \citep{Dalal:2006qt,Cutler:2009qv,Zhao:2010sz,Cai:2016sby,Cai:2017aea,Wang:2018lun,Zhang:2018byx,Zhang:2019ylr,Wang:2019tto,Zhao:2019gyk,Zhang:2019loq,Jin:2020hmc,Wang:2021srv,Yu:2023ico,Han:2023exn,Feng:2024lzh} and ``dark sirens'' \citep{Chen:2017rfc,Gray:2019ksv,Yu:2020vyy,Zhu:2021aat,Zhu:2021bpp,Song:2022siz,Jin:2023sfc,Jin:2023tou,Muttoni:2023prw,Li:2023gtu,Yun:2023ygz,Dong:2024bvw,Xiao:2024nmi,Zhu:2024qpp,Zheng:2024mbo}. Recently, \citet{Cao:2021zpf} proposed using GW bright sirens to calibrate distances in SGLs and to constrain $H_0$ and $\Omega_K$. Since only one bright siren, GW170817 \citep{LIGOScientific:2017adf}, has been observed, they implemented their method on mock data.

In this letter, we extend the method proposed by \citet{Cao:2021zpf} by incorporating GW dark sirens. Compared to GW bright sirens, GW dark sirens offer a broader redshift range and higher detection rates, primarily due to their independence from EM counterparts and the fact that their sources are typically more massive binary black holes (BBHs). By using GW dark sirens to calibrate SGLs, we can include more SGLTD data, thereby reducing statistical errors. Furthermore, by considering GW dark sirens, we are able, for the first time, to use real observed GW and SGLTD data to measure $H_0$ without relying on specific cosmological models.

\section{Methodology and data}

\subsection{The distance sum rule}
If space is homogeneous and isotropic, spacetime is described by the FLRW metric. Under the assumption that geometrical optics holds and the time-redshift relation $t(z)$ is monotonic, the angular diameter distance $D_{\rm A}(z_{\rm l},z_{\rm s})$ between a lens galaxy at redshift $z_{\rm l}$ and a source at redshift $z_{\rm s}$ can be expressed in terms of the dimensionless distance $d(z_{\rm l},z_{\rm s})=(1+z_{\rm s})H_0D_{\rm A}(z_{\rm l},z_{\rm s})$, defined as:
\begin{equation}
    \begin{aligned}
       & d(z_{\rm l},z_{\rm s})=\frac{1}{\sqrt{|\Omega_K|}}{\rm sinn}(\sqrt{|\Omega_K|}\int_{z_{\rm l}}^{z_{\rm s}}\frac{H_0}{H(z)}{\rm d}z),\\
&{\rm where}\\
        &\operatorname{sinn}(x)=\left\{\begin{array}{ll}
\sin (x), & \Omega_{K}<0, \\
x, & \Omega_{K}=0, \\
\sinh (x), & \Omega_{K}>0 .
\end{array}\right.
    \end{aligned}
\end{equation}
Here, $H(z)$ is the Hubble parameter, and $\Omega_K=-K/H_0^2$. $K$ is a constant describing the spatial curvature. For simplicity, we define $d_{\rm l}\equiv d(0,z_{\rm l})$, $d_{\rm s}\equiv d(0,z_{\rm s})$, and $d_{\rm ls}\equiv d(z_{\rm l},z_{\rm s})$, and these three dimensionless distances are connected via the distance sum rule \citep{Rasanen:2014mca}
\begin{equation}
    \begin{aligned}
        \frac{d_{\rm ls}}{d_{\rm s}}=\sqrt{1+\Omega_{K} d_{\rm l}^{2}}-\frac{d_{\rm l}}{d_{\rm s}} \sqrt{1+\Omega_{K} d_{\rm s}^{2}},
    \end{aligned}
\end{equation}
which can be further written as
\begin{equation}\label{eq:distance sum2}
    \begin{aligned}
        \frac{d_{\rm l}d_{\rm s}}{d_{\rm ls}}=\frac{1}{\sqrt{1 / d_{\rm l}^{2}+\Omega_{K}}-\sqrt{1 / d_{\rm s}^{2}+\Omega_{K}}}.
    \end{aligned}
\end{equation}

\subsection{Time delay and distances}
In time-delay system, the arrival time delay $\Delta t_{ij}$ between two images at ${\bm\theta}_i$ and ${\bm\theta}_j$ is given by \citep{Refsdal:1964blz,Suyu:2009by}
\begin{equation}
    \begin{aligned}\label{eq:time delay}
       \Delta t_{ij}=D_{\Delta t}\Delta \phi_{i,j}=\frac{1}{H_0}\frac{d_{\rm l}d_{\rm s}}{d_{\rm ls}}\Delta\phi_{ij},
    \end{aligned}
\end{equation}
where $D_{\Delta t}$ is the time-delay distance, and $\Delta\phi_{ij}$ is the Fermat potential difference of these two images. Given $\Delta t_{i,j}$, $d_{\rm l}$, $d_{\rm s}$, and an inference of $\Delta \phi_{i,j}$, we can determine $H_0$ and $\Omega_K$ from Equations~(\ref{eq:distance sum2}) and (\ref{eq:time delay}) without relying on a specific cosmological model.

\subsection{Obtaining redshifts for GW standard sirens}
Detections of EM counterparts can help localize GW sources' host galaxies, thus determining redshifts \citep{Schutz:1986gp,LIGOScientific:2017adf,LIGOScientific:2017vwq,LIGOScientific:2017zic}. We adopt two methods to obtain redshifts for GW dark sirens without EM counterparts. The first method uses the population model of GW sources, called the spectral siren method \citep{Taylor:2011fs,Ezquiaga:2022zkx}. The waveform phase of the GW standard siren determines the redshifted masses $m_{1,z}$ and $m_{2,z}$, which are related to the source-frame masses $m_1$ and $m_2$ via
\begin{equation}\label{eq:redshifted mass}
    m_i=\frac{m_{i,z}}{1+z}.
\end{equation}
We can break the degeneracy between masses and redshifts by modeling the mass and redshift distributions in the source frame, hence extracting redshift.

The second method involves identifying potential host galaxies of GW sources using galaxy catalogs, known as the dark siren approach \citep{Schutz:1986gp,Mastrogiovanni:2023emh,Gray:2023wgj}. The GW observation provides a coarse localization of the GW source, while the galaxy catalog supplies redshift information for galaxies within this region, which may include the host galaxy. By assuming that the probability of a galaxy being the host is related to its luminosity, the galaxies in the localization region can be used to infer an approximate redshift distribution for the GW source. Due to the limited observational capabilities of survey telescopes, the completeness of galaxy catalogs decreases with distance. In regions where the catalog is incomplete, redshift information is supplemented through the spectral siren method using GW population models.

\begin{figure*}[!htbp]
    \centering
    \includegraphics[width=0.3\textwidth]{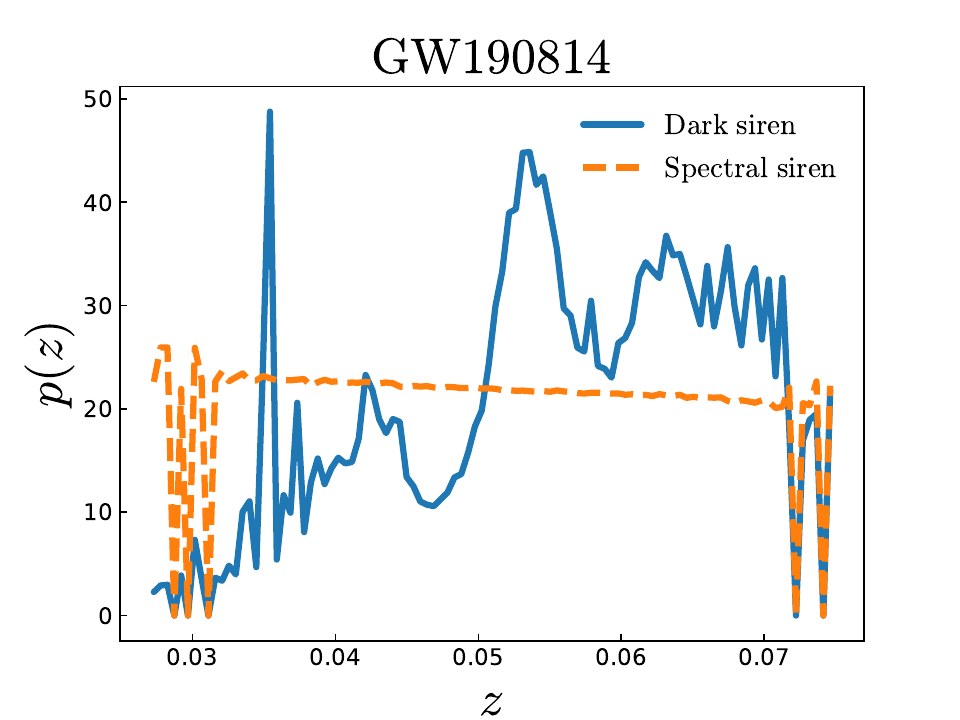}
    \includegraphics[width=0.3\textwidth]{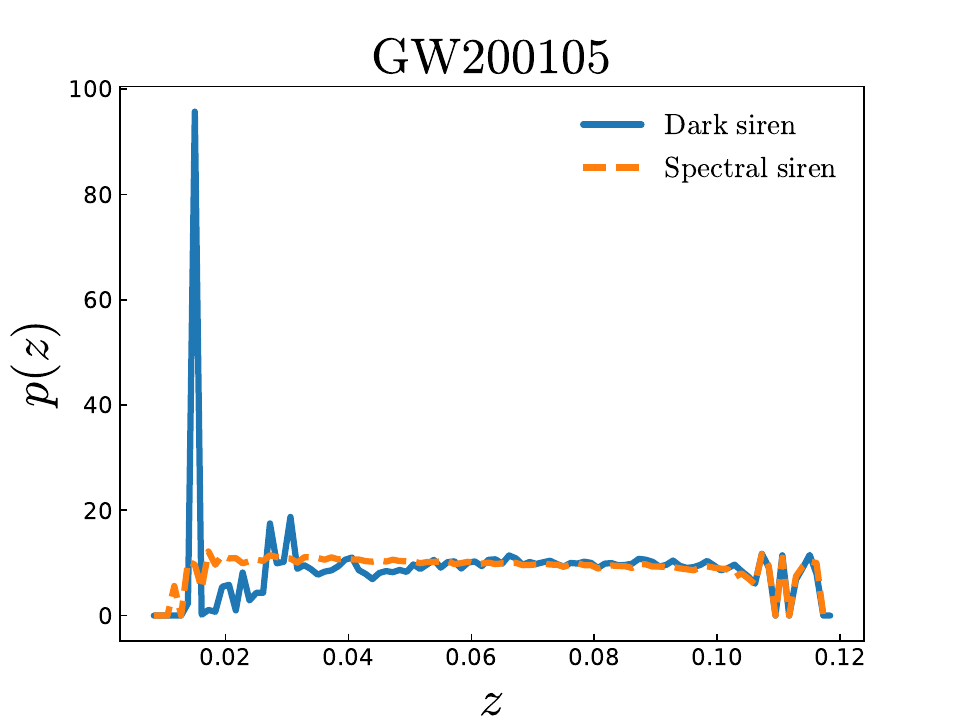}
    \includegraphics[width=0.3\textwidth]{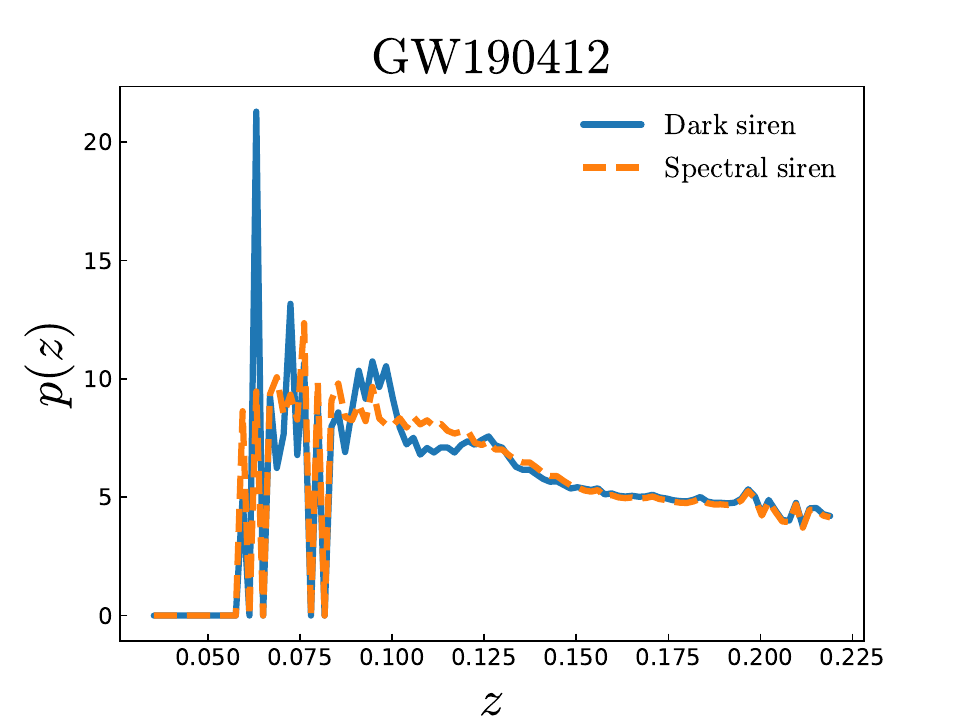}
    \caption{Redshift distributions for GW190814, GW200105, and GW190412 obtained using different methods, with cosmological and population model parameters fixed.}
    \label{fig:compare}
\end{figure*}

In Figure~\ref{fig:compare}, we compare two methods for estimating redshifts of GW events without EM counterparts, using GW190814, GW200105, and GW190412 as illustrative examples. The dark siren approach enhances redshift inference by incorporating galaxy catalogs. Its effectiveness depends on the completeness of the galaxy catalog within the GW event's localization region and the accuracy of the localization. As shown in the figure, GW190814 is a low-redshift event with excellent localization, making the dark siren approach more effective than the spectral siren method. GW200105 has a similar redshift to GW190814 but poorer localization, so the inclusion of the galaxy catalog does not noticeably improve the redshift estimate. GW190412 is the second best located GW event after GW190814 but at a higher redshift, where galaxy catalogs are less complete, resulting in limited improvement. For specific redshifts and localizations of above GW events, refer to \citet{LIGOScientific:2021aug}.

\subsection{Gravitational wave data}
We use 47 GW standard sirens with a signal-to-noise ratio greater than 11 and an inverse false alarm rate exceeding 4 yr from the third Gravitational-Wave Transient Catalog (GWTC-3), covering a redshift range of approximately $z < 0.8$. The sample consists of 42 BBHs, three neutron star-black holes (NSBHs), and two binary neutron stars (BNSs).

We obtain the redshift of GW170817 from its EM counterpart \citep{LIGOScientific:2017zic}. For dark sirens, when applying the spectral siren method, we consider only 42 BBH events; when employing the dark siren method, we include all dark sirens and incorporate $K$-band galaxies from the extended galaxy list for the advanced detector era (GLADE+) \citep{Dalya:2021ewn}. We assume that the absolute magnitudes of $K$-band galaxies in GLADE+ follow the Schechter function \citep{Kochanek:2000im}.

Following \citet{LIGOScientific:2021aug}, we consider three BBH mass distribution models: Power Law + Peak, Broken Power Law, and Truncated. The Power Law + Peak and Broken Power Law models are relatively simple, each characterized by a single structure in the power law, and remain preferred by the GWTC-3 data \citep{KAGRA:2021duu}. The Truncated model is strongly disfavored by GWTC-3 data, but we include it for comparison. We assume a uniform distribution between 1 $M_{\odot}$ and 3 $M_{\odot}$ for the NS mass model. For NSBH, the black hole mass distribution is the same as the primary black hole in BBH, and the neutron star mass distribution is the same as that in BNS. We assume that the binary formation rate follows the star formation rate and model the redshift distribution of GW sources using a phenomenological approach from \citet{Madau:2014bja}. Consistent with \citet{LIGOScientific:2021aug}, we neglect the potential evolution of the BBH mass distribution with redshift and the selection effect of the spin distribution. The expected evolution of the BBH mass distribution is below the current statistical uncertainties in the redshift range of GW data \citep{Fishbach:2021mhp,vanSon:2021zpk}, and including the spin distribution does not alter the detection probability by more than a factor of two \citep{Ng:2018neg}.

\subsection{Strong lensing data}
We consider only RXJ1131-1231 \citep{Suyu:2012aa,Suyu:2013kha,H0LiCOW:2019xdh} in our analysis, as it is entirely within the redshift coverage of our GW data. Specifically, the spectroscopic redshifts of the lens and the source are {$z_{\rm l}=0.295$} \citep{Sluse:2003iy} and $z_{\rm s}=0.657$ \citep{Sluse:2007cn}. The mass-sheet transform (MST) is a major source of systematic error in the time-delay cosmology, as it preserves lensing observables but rescales the absolute time delay and the inferred value of $H_0$. The H0LiCOW collaboration addressed MST by modeling the lens galaxy’s mass density profile using a composite model that combines a singular elliptical power-law model with a baryon + Navarro-Frenk-White (NFW) \citep{Navarro:1996gj} dark matter halo model, with model selection weighted by the Bayesian information criterion \citep{H0LiCOW:2019pvv}.
In contrast, the TDCOSMO collaboration employed a hierarchical Bayesian analysis to break MST using only stellar kinematics \citep{Birrer:2020tax}. We employ both the H0LiCOW and TDCOSMO methods and evaluate their impact on our results.

\begin{table*}
\centering
\caption{Constraint values and precisions of $H_0$ with various data choices, BBH mass distribution models, and MST treatment methods. Most of the results are obtained using the third-order polynomial model. For comparison, four cases based on the $\Lambda$CDM model are also considered, and are labeled as $\Lambda$CDM in the ``Condition'' column.
$H_0$ is in units of $\rm{km}~\rm{s}^{-1}~\rm{Mpc}^{-1}$, and $\Delta H_0/H_0$ represents the precisions of $H_0$ constraints at the 68\% confidence level. }\label{tab:constraint results}
\centering
\renewcommand{\arraystretch}{2}
\begin{tabular}{ccccc}
\hline\hline
\makebox[0.19\textwidth][c]{Data} & \makebox[0.19\textwidth][c]{BBH's mass model} & \makebox[0.19\textwidth][c]{MST treatment method} & \makebox[0.19\textwidth][c]{$H_0$ ($\Delta H_0/H_0$)} & \makebox[0.15\textwidth][c]{Condition} \\
\hline
\multirow{5}{*}{47 events + RXJ1131-1231}            & Power Law + Peak                              & H0LiCOW                                           & $73.22^{+5.95}_{-5.43}$ (7.77\%)              & none\\
                                                     & Power Law + Peak                              & H0LiCOW                                           & $76.26^{+10.60}_{-7.34}$ (11.76\%)            & unfix $\Omega_K$\\
                                                     & Power Law + Peak                              & TDCOSMO                                           & $70.40^{+8.03}_{-5.60}$ (9.68\%)              & none\\
                                                     & Power Law + Peak                              & H0LiCOW                                           & $76.94^{+3.34}_{-3.24}$ (4.28\%)              & $\Lambda$CDM\\
                                                     & Power Law + Peak                              & TDCOSMO                                           & $70.54^{+6.41}_{-5.35}$ (8.34\%)              & $\Lambda$CDM\\ 
\hline
\multirow{4}{*}{42 events + RXJ1131-1231}            & Power Law + Peak                              & H0LiCOW                                           & $73.78^{+6.48}_{-6.85}$ (9.03\%)              & without GLADE+ \\
                                                     & Broken Power Law                              & H0LiCOW                                           & $74.60^{+6.40}_{-6.95}$ (8.95\%)              & without GLADE+ \\
                                                     & Truncated                                     & H0LiCOW                                           & $74.98^{+6.35}_{-6.95}$ (8.87\%)              & without GLADE+ \\
                                                     & Power Law + Peak                              & H0LiCOW                                           & $73.28^{+6.16}_{-6.24}$ (8.46\%)              & none\\
\hline
43 events + RXJ1131-1231                             & Power Law + Peak                              & H0LiCOW                                           & $73.27^{+5.90}_{-5.64}$ (7.87\%)              & without GLADE+ \\
\hline
\multirow{2}{*}{RXJ1131-1231}                        & none                                          & H0LiCOW                                           & $77.33^{+3.34}_{-3.30}$ (4.29\%)              & $\Lambda$CDM\\
                                                     & none                                          & TDCOSMO                                           & $68.65^{+8.15}_{-7.01}$ (11.04\%)             & $\Lambda$CDM\\  
\hline\hline
\end{tabular}
\end{table*}

\section{Results and discussions}

We constrain $d(z)$ and $H_0$ through the public Bayesian analysis code \texttt{Bilby} \citep{Ashton:2018jfp} and the Markov chain Monte Carlo sampler \texttt{emcee} \citep{Foreman-Mackey:2012any}, using GW data and SGLTD data. The logarithm of the total likelihood $\mathcal{L}$ is given by
\begin{equation}
    {\rm ln}(\mathcal{L}) = {\rm ln}(\mathcal{L}_{\rm SGL})+{\rm ln}(\mathcal{L}_{\rm GW}),
\end{equation}
where $\mathcal{L}_{\rm SGL}$ is the likelihood function of SGLTDs. When using the H0LiCOW method to break MST, we obtain $\mathcal{L}_{\rm SGL}$ through kernel density estimation based on $D_{\Delta t}$ and $D_{\rm A}(z_{\rm l})$ samples from \citet{H0LiCOW:2019xdh}. When applying the TDCOSMO method to break MST, we use the public code from \citet{Birrer:2020tax} to calculate $\mathcal{L}_{\rm SGL}$. We incorporate 33 SGLs from the Sloan Lens ACS survey \citep{Bolton:2005nf} to further constrain the lens galaxy's mass density profiles and jointly infer both cosmological and SGL model parameters. $\mathcal{L}_{\rm GW}$ is the likelihood function of the GW data. We use the public code \texttt{ICAROGW} \citep{Mastrogiovanni:2023zbw} to calculate $\mathcal{L}_{\rm GW}$.

Following \citet{Collett:2019hrr,Cao:2021zpf}, we model $d(z)$ with a polynomial. By calculating the Bayesian information criterion for fitting GW data, we find that a third-order polynomial is sufficient to fit current GW data. We assume $d(0)=0$ and $d'(0)=1$, so the third-order polynomial has two free parameters.

We adopt flat priors for all free parameters, with prior ranges specified as follows: $H_0\in[20,140]$ $\rm{km}~\rm{s}^{-1}~\rm{Mpc}^{-1}$ and the $i$-order coefficient $c_i\in [-1,1]$ in polynomials. Additionally, we consider cases of assuming the $\Lambda$CDM model with $\Omega_{\rm m} \in [0, 1]$ for comparison. The model parameters of GW and SGLTD data in the hierarchical likelihood may degenerate with $H_0$, so they appear as free parameters, with prior ranges taken from \citet{LIGOScientific:2021aug} and \citet{Birrer:2020tax}, respectively. In this work, we obtain $H_0$ constraint results assuming $\Omega_K=0$ in most cases. We also consider releasing $\Omega_K$ as a free parameter with the prior range of $\Omega_K\in[-2,2]$.

\begin{figure}
    \centering
    \includegraphics[width=0.45\textwidth]{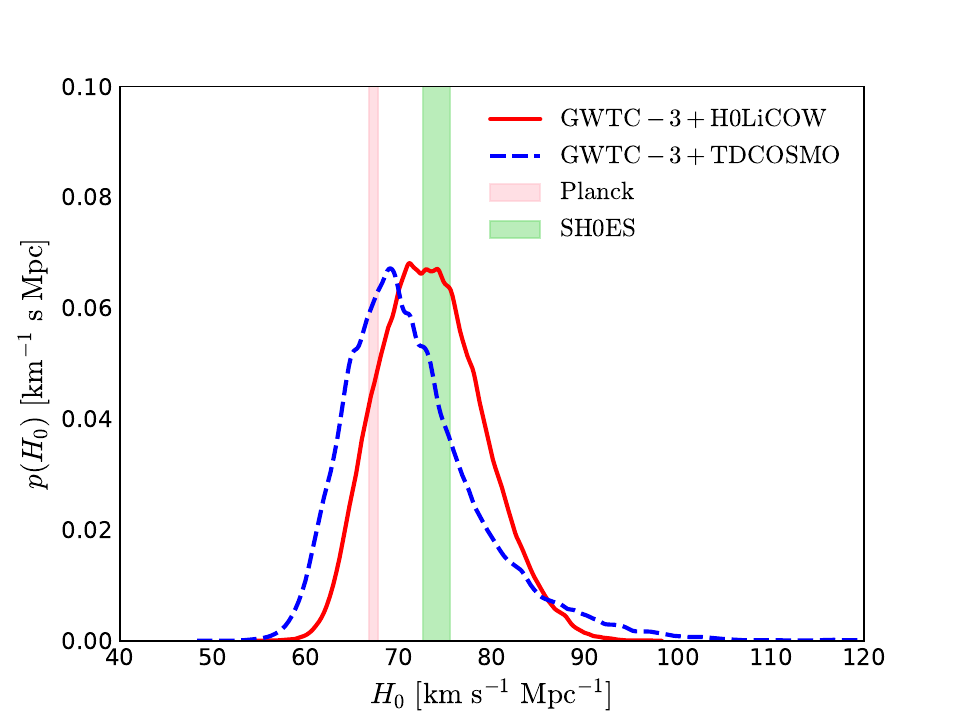}
    \caption{Posterior distributions of $H_0$ using 47 GW events and RXJ1131-1231 in the polynomial model. The red solid line and the blue dashed line represent the cases of using the H0LiCOW method and the TDCOSMO method to break MST, respectively. The pink and green shaded areas show the 68\% confidence interval from the Planck 2018 CMB inference in the $\Lambda$CDM model and the SH0ES distance ladder measurement in the local universe, respectively.
    }\label{fig:H0}
\end{figure}

We jointly constrain both cosmological and population model parameters using GW and SGLTD data, with the $H_0$ constraints summarized in Table~\ref{tab:constraint results}. We consider a range of observational data combinations and model assumptions to comprehensively assess their impacts on our results.

We obtain our baseline result by combining 47 GW events with RXJ1131-1231, as shown in the first row of Table~\ref{tab:constraint results}. In this analysis, we incorporate the GLADE+ catalog to determine the redshifts of GW dark sirens and apply mass model assumptions to break MST.
We obtain $H_0 = 73.22^{+5.95}_{-5.43}$ $\rm{km}~\rm{s}^{-1}~\rm{Mpc}^{-1}$, which is more consistent with SH0ES distance ladder measurements \citep{Riess:2021jrx} and differs from the Planck TT, TE, EE+low E result \citep{Planck:2018vyg} in the $\Lambda$CDM model by 5.95 $\rm{km}~\rm{s}^{-1}~\rm{Mpc}^{-1}$, corresponding to a deviation of approximately 1$\sigma$.

When allowing $\Omega_K$ to vary and using the same data and model assumptions as in the baseline case, we obtain $H_0 = 76.26^{+10.60}_{-7.34}$ km s$^{-1}$ Mpc$^{-1}$ and $\Omega_K = 0.79^{+0.83}_{-1.24}$, with the $H_0$ result presented in the second row of Table~\ref{tab:constraint results}. Compared with the baseline result, this measurement yields a higher central value for $H_0$ and a reduced constraint precision. The deviation in $H_0$'s central value after unfixing $\Omega_K$ is consistent with \citet{Collett:2019hrr}, due to the positive correlation between $H_0$ and $\Omega_K$. Our $\Omega_K$ result favors an open universe, in agreement with other late-time cosmological probes \citep{Wu:2024faw}, and deviates from the Planck CMB result \citep{Planck:2018vyg}, though the difference is less than 1$\sigma$ due to the poor constraint precision.

By applying the hierarchical inference approach to break MST, we obtain $H_0 = 70.40^{+8.03}_{-5.60}$ $\rm{km}~\rm{s}^{-1}~\rm{Mpc}^{-1}$, as shown in the third row of Table~\ref{tab:constraint results}. Compared with the baseline result, this approach estimates a lower central value for $H_0$ with a larger uncertainty. We present the $H_0$ posteriors using different MST treatment methods, set against the backdrop of the Hubble tension in Figure~\ref{fig:H0}. To further investigate the deviation in the central value of $H_0$, we constrain $H_0$ using only SGLTD in the $\Lambda$CDM model, with the results shown in the last two rows of Table~\ref{tab:constraint results}. We observe a similar bias in the central value, but with a larger magnitude. The discrepancy between different MST treatment methods may arise from the use of a single SGLTD dataset, which could have large statistical errors. Furthermore, we incorporate a set of SLACS lenses to further constrain the MST when applying the TDCOSMO method. Nevertheless, the results obtained using different MST treatment methods are in statistical agreement with each other. The worsening of $H_0$ constraint precision using the TDCOSMO method is expected, as the hierarchical approach breaks MST using only kinematic information from the lens galaxy, and the uncertainties propagate into the cosmological parameter inference \citep{Birrer:2015fsm,Birrer:2020tax}.

To ensure a fair comparison between the polynomial and $\Lambda$CDM models, we also constrain $H_0$ using 47 GW events in combination with RXJ1131–1231, as shown in the fourth and fifth rows of Table~\ref{tab:constraint results}. We find that the $H_0$ values inferred using either the H0LiCOW or TDCOSMO method under the polynomial model are statistically consistent with those obtained under the $\Lambda$CDM model, with differences within $1\sigma$. Furthermore, compared with results using only the SGLTD data, incorporating GW data induces only minor changes in the inferred $H_0$, particularly for the H0LiCOW method, indicating that the SGLTD data dominate the constraint.

In the GW data, the determination of redshifts for dark sirens relies on GW population models. Since BBH events constitute the majority of current GW data, their population model assumptions may influence the results. To evaluate this, we constrain $H_0$ using 42 BBHs and RXJ1131-1231, testing three BBH mass models: Power Law + Peak, Broken Power Law, and Truncated. The results are presented in the sixth to eighth rows of Table~\ref{tab:constraint results}. To isolate the impact of the GW population model, we exclude the GLADE+ catalog and use the spectral siren method to determine the redshifts of the 42 BBH events. We find that the estimated central values of $H_0$ from the three BBH mass models are nearly consistent, which differs from the results obtained using 42 BBHs alone in the $w$CDM model \citep{LIGOScientific:2021aug}, particularly when assuming the Truncated model.

Galaxy catalogs and EM counterparts are two types of EM observations incorporated in the GW data. To evaluate the impact of including GLADE+ on our results, we constrain $H_0$ using 42 BBH events and RXJ1131-1231, incorporating GLADE+, as shown in the ninth row of Table~\ref{tab:constraint results}. We observe only a 0.57\% improvement in the $H_0$ constraint precision compared with the result obtained using the same data and model assumptions, but without GLADE+. This is primarily due to the limited completeness of GLADE+ in the $K$ band, which decreases with increasing distance, reaching approximately 50\% at around 420 Mpc \citep{Dalya:2021ewn}. Furthermore, we observe a $\sim40$\% improvement in the $H_0$ constraint precision compared with the result using the bright siren GW170817 in the Hubble law \citep{LIGOScientific:2017adf}. When we combine GW170817 with the 42 BBH events and RXJ1131-1231, we achieve a similar $H_0$ constraint precision as the baseline result, as shown in the tenth row of Table~\ref{tab:constraint results}.

\section{Conclusions}
The Hubble tension has emerged as a crisis in cosmology, and a late-time, cosmological-model-independent determination of $H_0$ could help resolve this issue. By using GW standard sirens to calibrate distances in SGLs, a cosmological-model-independent measurement of $H_0$ can be achieved in the FLRW metric through the distance-sum rule. Furthermore, GW standard sirens provide self-calibrated luminosity distance measurements and can offer more reliable calibrations for distances in SGLs than other distance indicators, such as SNe Ia and quasars. In this letter, we propose using GW dark sirens to extend the limited redshift range of the GW bright siren and, for the first time, use 47 GW events from GWTC-3 and RXJ1131-1231 to measure $H_0$ without assuming a specific cosmological model.

By fixing $\Omega_K=0$ and breaking MST through the lens galaxy's mass model, we obtain $H_0 = 73.22^{+5.95}_{-5.43}$ $\rm{km}~\rm{s}^{-1}~\rm{Mpc}^{-1}$ with a precision of 7.77\%. This result deviates by approximately $1\sigma$ from the Planck 2018 CMB results in the $\Lambda$CDM model. When using hierarchical inference to break MST with stellar kinematics, we obtain $H_0 = 70.40^{+8.03}_{-5.60}$ $\rm{km}~\rm{s}^{-1}~\rm{Mpc}^{-1}$ with a precision of 9.68\%. Upon unfixing $\Omega_K$, we obtain $H_0 = 76.26^{+10.60}_{-7.34}$ $\rm{km}~\rm{s}^{-1}~\rm{Mpc}^{-1}$ with a precision of 11.76\%, and $\Omega_K = 0.79^{+0.83}_{-1.24}$, using the lens galaxy's mass model to break MST.

We find that the method used to break MST affects our results; however, different approaches yield statistically consistent outcomes. The choice of BBH mass models becomes less significant compared with analyses using GW data alone. The impact of the GLADE+ catalog is minimal, while incorporating GW170817 and its EM counterpart notably improves the $H_0$ constraint precision. Remarkably, the $H_0$ constraint is primarily driven by the SGLTD data, indicating that future improvements could rely on incorporating more SGLTDs. In this context, GW dark sirens have the advantage of being at higher redshifts, which can calibrate more SGL systems. Currently, combining 42 BBH events with RXJ1131-1231 improves the $H_0$ constraint precision by approximately 40\% compared with that obtained from GW170817 using the Hubble law \citep{LIGOScientific:2017adf}. In the future, observing more high-redshift GW dark sirens and using them to calibrate additional SGLTDs could significantly reduce statistical uncertainties, enabling precise, cosmological-model-independent $H_0$ measurements.

\section*{Acknowledgments}

We are grateful to Ji-Guo Zhang, Guo-Hong Du, and Xianzhe TZ Tang for fruitful discussions. This work was supported by the National SKA Program of China (Grants Nos. 2022SKA0110200 and 2022SKA0110203), the National Natural Science Foundation of China (Grants Nos. 12473001, 11975072, 11835009, 11875102, and 12205039), the China Manned Space Program (Grant No. CMS-CSST-2025-A02), and the 111 Project (Grant No. B16009).

\bibliography{ref}

\begin{thebibliography}{}
\expandafter\ifx\csname natexlab\endcsname\relax\def\natexlab#1{#1}\fi
\providecommand{\url}[1]{\href{#1}{#1}}
\providecommand{\dodoi}[1]{doi:~\href{http://doi.org/#1}{\nolinkurl{#1}}}
\providecommand{\doeprint}[1]{\href{http://ascl.net/#1}{\nolinkurl{http://ascl.net/#1}}}
\providecommand{\doarXiv}[1]{\href{https://arxiv.org/abs/#1}{\nolinkurl{https://arxiv.org/abs/#1}}}

\bibitem[{Abbott {et~al.}(2017{\natexlab{a}})}]{LIGOScientific:2017adf}
Abbott, B.~P., {et~al.} 2017{\natexlab{a}}, Nature, 551, 85,
  \dodoi{10.1038/nature24471}

\bibitem[{Abbott {et~al.}(2017{\natexlab{b}})}]{LIGOScientific:2017vwq}
---. 2017{\natexlab{b}}, Phys. Rev. Lett., 119, 161101,
  \dodoi{10.1103/PhysRevLett.119.161101}

\bibitem[{Abbott {et~al.}(2017{\natexlab{c}})}]{LIGOScientific:2017zic}
---. 2017{\natexlab{c}}, Astrophys. J. Lett., 848, L13,
  \dodoi{10.3847/2041-8213/aa920c}

\bibitem[{Abbott {et~al.}(2023{\natexlab{a}})}]{LIGOScientific:2021aug}
Abbott, R., {et~al.} 2023{\natexlab{a}}, Astrophys. J., 949, 76,
  \dodoi{10.3847/1538-4357/ac74bb}

\bibitem[{Abbott {et~al.}(2023{\natexlab{b}})}]{KAGRA:2021duu}
---. 2023{\natexlab{b}}, Phys. Rev. X, 13, 011048,
  \dodoi{10.1103/PhysRevX.13.011048}

\bibitem[{Aghanim {et~al.}(2017)}]{Planck:2016tof}
Aghanim, N., {et~al.} 2017, Astron. Astrophys., 607, A95,
  \dodoi{10.1051/0004-6361/201629504}

\bibitem[{Aghanim {et~al.}(2020)}]{Planck:2018vyg}
---. 2020, Astron. Astrophys., 641, A6, \dodoi{10.1051/0004-6361/201833910}

\bibitem[{Akrami {et~al.}(2020)}]{Planck:2019evm}
Akrami, Y., {et~al.} 2020, Astron. Astrophys., 641, A7,
  \dodoi{10.1051/0004-6361/201935201}

\bibitem[{Ashton {et~al.}(2019)}]{Ashton:2018jfp}
Ashton, G., {et~al.} 2019, Astrophys. J. Suppl., 241, 27,
  \dodoi{10.3847/1538-4365/ab06fc}

\bibitem[{Birrer {et~al.}(2016)Birrer, Amara, \& Refregier}]{Birrer:2015fsm}
Birrer, S., Amara, A., \& Refregier, A. 2016, JCAP, 08, 020,
  \dodoi{10.1088/1475-7516/2016/08/020}

\bibitem[{Birrer {et~al.}(2020)}]{Birrer:2020tax}
Birrer, S., {et~al.} 2020, Astron. Astrophys., 643, A165,
  \dodoi{10.1051/0004-6361/202038861}

\bibitem[{Bolton {et~al.}(2006)Bolton, Burles, Koopmans, Treu, \&
  Moustakas}]{Bolton:2005nf}
Bolton, A.~S., Burles, S., Koopmans, L. V.~E., Treu, T., \& Moustakas, L.~A.
  2006, Astrophys. J., 638, 703, \dodoi{10.1086/498884}

\bibitem[{Cai {et~al.}(2018)Cai, Liu, Liu, Wang, \& Yang}]{Cai:2017aea}
Cai, R.-G., Liu, T.-B., Liu, X.-W., Wang, S.-J., \& Yang, T. 2018, Phys. Rev.
  D, 97, 103005, \dodoi{10.1103/PhysRevD.97.103005}

\bibitem[{Cai \& Yang(2017)}]{Cai:2016sby}
Cai, R.-G., \& Yang, T. 2017, Phys. Rev. D, 95, 044024,
  \dodoi{10.1103/PhysRevD.95.044024}

\bibitem[{Cao {et~al.}(2022)Cao, Zheng, Qi, Zhang, \& Zhu}]{Cao:2021zpf}
Cao, M.-D., Zheng, J., Qi, J.-Z., Zhang, X., \& Zhu, Z.-H. 2022, Astrophys. J.,
  934, 108, \dodoi{10.3847/1538-4357/ac7ce4}

\bibitem[{Castello {et~al.}(2022)Castello, H\"og\r{a}s, \&
  M\"ortsell}]{Castello:2021uad}
Castello, S., H\"og\r{a}s, M., \& M\"ortsell, E. 2022, JCAP, 07, 003,
  \dodoi{10.1088/1475-7516/2022/07/003}

\bibitem[{Chen {et~al.}(2019)}]{H0LiCOW:2019xdh}
Chen, G. C.~F., {et~al.} 2019, Mon. Not. Roy. Astron. Soc., 490, 1743,
  \dodoi{10.1093/mnras/stz2547}

\bibitem[{Chen {et~al.}(2018)Chen, Fishbach, \& Holz}]{Chen:2017rfc}
Chen, H.-Y., Fishbach, M., \& Holz, D.~E. 2018, Nature, 562, 545,
  \dodoi{10.1038/s41586-018-0606-0}

\bibitem[{Collett {et~al.}(2019)Collett, Montanari, \&
  Rasanen}]{Collett:2019hrr}
Collett, T., Montanari, F., \& Rasanen, S. 2019, Phys. Rev. Lett., 123, 231101,
  \dodoi{10.1103/PhysRevLett.123.231101}

\bibitem[{Cutler \& Holz(2009)}]{Cutler:2009qv}
Cutler, C., \& Holz, D.~E. 2009, Phys. Rev. D, 80, 104009,
  \dodoi{10.1103/PhysRevD.80.104009}

\bibitem[{Dalal {et~al.}(2006)Dalal, Holz, Hughes, \& Jain}]{Dalal:2006qt}
Dalal, N., Holz, D.~E., Hughes, S.~A., \& Jain, B. 2006, Phys. Rev. D, 74,
  063006, \dodoi{10.1103/PhysRevD.74.063006}

\bibitem[{D\'alya {et~al.}(2022)}]{Dalya:2021ewn}
D\'alya, G., {et~al.} 2022, Mon. Not. Roy. Astron. Soc., 514, 1403,
  \dodoi{10.1093/mnras/stac1443}

\bibitem[{D'Amico {et~al.}(2021)D'Amico, Senatore, Zhang, \&
  Zheng}]{DAmico:2020ods}
D'Amico, G., Senatore, L., Zhang, P., \& Zheng, H. 2021, JCAP, 05, 072,
  \dodoi{10.1088/1475-7516/2021/05/072}

\bibitem[{Dong {et~al.}(2024)Dong, Song, Jin, Zhang, \& Zhang}]{Dong:2024bvw}
Dong, Y.-Y., Song, J.-Y., Jin, S.-J., Zhang, J.-F., \& Zhang, X. 2024.
\newblock \doarXiv{2404.18188}

\bibitem[{Ezquiaga \& Holz(2022)}]{Ezquiaga:2022zkx}
Ezquiaga, J.~M., \& Holz, D.~E. 2022, Phys. Rev. Lett., 129, 061102,
  \dodoi{10.1103/PhysRevLett.129.061102}

\bibitem[{Feng {et~al.}(2024)Feng, Han, Zhang, \& Zhang}]{Feng:2024lzh}
Feng, L., Han, T., Zhang, J.-F., \& Zhang, X. 2024, Chin. Phys. C, 48, 095104,
  \dodoi{10.1088/1674-1137/ad5ae4}

\bibitem[{Fishbach \& Kalogera(2021)}]{Fishbach:2021mhp}
Fishbach, M., \& Kalogera, V. 2021, Astrophys. J. Lett., 914, L30,
  \dodoi{10.3847/2041-8213/ac05c4}

\bibitem[{Foreman-Mackey {et~al.}(2013)Foreman-Mackey, Hogg, Lang, \&
  Goodman}]{Foreman-Mackey:2012any}
Foreman-Mackey, D., Hogg, D.~W., Lang, D., \& Goodman, J. 2013, Publ. Astron.
  Soc. Pac., 125, 306, \dodoi{10.1086/670067}

\bibitem[{Freedman {et~al.}(2024)Freedman, Madore, Jang, Hoyt, Lee, \&
  Owens}]{Freedman:2024eph}
Freedman, W.~L., Madore, B.~F., Jang, I.~S., {et~al.} 2024.
\newblock \doarXiv{2408.06153}

\bibitem[{Gray {et~al.}(2020)}]{Gray:2019ksv}
Gray, R., {et~al.} 2020, Phys. Rev. D, 101, 122001,
  \dodoi{10.1103/PhysRevD.101.122001}

\bibitem[{Gray {et~al.}(2023)}]{Gray:2023wgj}
---. 2023, JCAP, 12, 023, \dodoi{10.1088/1475-7516/2023/12/023}

\bibitem[{Guo {et~al.}(2019)Guo, Zhang, \& Zhang}]{Guo:2018ans}
Guo, R.-Y., Zhang, J.-F., \& Zhang, X. 2019, JCAP, 02, 054,
  \dodoi{10.1088/1475-7516/2019/02/054}

\bibitem[{Han {et~al.}(2024)Han, Jin, Zhang, \& Zhang}]{Han:2023exn}
Han, T., Jin, S.-J., Zhang, J.-F., \& Zhang, X. 2024, Eur. Phys. J. C, 84, 663,
  \dodoi{10.1140/epjc/s10052-024-12999-w}

\bibitem[{Hill {et~al.}(2020)Hill, McDonough, Toomey, \&
  Alexander}]{Hill:2020osr}
Hill, J.~C., McDonough, E., Toomey, M.~W., \& Alexander, S. 2020, Phys. Rev. D,
  102, 043507, \dodoi{10.1103/PhysRevD.102.043507}

\bibitem[{H\"og\r{a}s \& M\"ortsell(2024)}]{Hogas:2024qlt}
H\"og\r{a}s, M., \& M\"ortsell, E. 2024.
\newblock \doarXiv{2412.07840}

\bibitem[{Huang {et~al.}(2024)Huang, Cai, Wang, Liu, \& Yao}]{Huang:2024gfw}
Huang, L., Cai, R.-G., Wang, S.-J., Liu, J.-Q., \& Yao, Y.-H. 2024.
\newblock \doarXiv{2410.06053}

\bibitem[{Huang {et~al.}(2025)Huang, Wang, \& Yu}]{Huang:2024erq}
Huang, L., Wang, S.-J., \& Yu, W.-W. 2025, Sci. China Phys. Mech. Astron., 68,
  220413, \dodoi{10.1007/s11433-024-2528-8}

\bibitem[{Jin {et~al.}(2020)Jin, He, Xu, Zhang, \& Zhang}]{Jin:2020hmc}
Jin, S.-J., He, D.-Z., Xu, Y., Zhang, J.-F., \& Zhang, X. 2020, JCAP, 03, 051,
  \dodoi{10.1088/1475-7516/2020/03/051}

\bibitem[{Jin {et~al.}(2024{\natexlab{a}})Jin, Zhang, Song, Zhang, \&
  Zhang}]{Jin:2023sfc}
Jin, S.-J., Zhang, Y.-Z., Song, J.-Y., Zhang, J.-F., \& Zhang, X.
  2024{\natexlab{a}}, Sci. China Phys. Mech. Astron., 67, 220412,
  \dodoi{10.1007/s11433-023-2276-1}

\bibitem[{Jin {et~al.}(2024{\natexlab{b}})Jin, Zhu, Song, Han, Zhang, \&
  Zhang}]{Jin:2023tou}
Jin, S.-J., Zhu, R.-Q., Song, J.-Y., {et~al.} 2024{\natexlab{b}}, JCAP, 08,
  050, \dodoi{10.1088/1475-7516/2024/08/050}

\bibitem[{Jones {et~al.}(2018)}]{Jones:2018vbn}
Jones, D.~O., {et~al.} 2018, Astrophys. J., 867, 108,
  \dodoi{10.3847/1538-4357/aae2b9}

\bibitem[{Kenworthy {et~al.}(2019)Kenworthy, Scolnic, \&
  Riess}]{Kenworthy:2019qwq}
Kenworthy, W.~D., Scolnic, D., \& Riess, A. 2019, Astrophys. J., 875, 145,
  \dodoi{10.3847/1538-4357/ab0ebf}

\bibitem[{Kochanek {et~al.}(2001)Kochanek, Pahre, Falco, Huchra, Mader,
  Jarrett, Chester, Cutri, \& Schneider}]{Kochanek:2000im}
Kochanek, C.~S., Pahre, M.~A., Falco, E.~E., {et~al.} 2001, Astrophys. J., 560,
  566, \dodoi{10.1086/322488}

\bibitem[{Lee {et~al.}(2024)Lee, Freedman, Madore, Jang, Owens, \&
  Hoyt}]{Lee:2024qzr}
Lee, A.~J., Freedman, W.~L., Madore, B.~F., {et~al.} 2024.
\newblock \doarXiv{2408.03474}

\bibitem[{Li {et~al.}(2024{\natexlab{a}})Li, Riess, Casertano, Anand, Scolnic,
  Yuan, Breuval, \& Huang}]{Li:2024yoe}
Li, S., Riess, A.~G., Casertano, S., {et~al.} 2024{\natexlab{a}}, Astrophys.
  J., 966, 20, \dodoi{10.3847/1538-4357/ad2f2b}

\bibitem[{Li {et~al.}(2024{\natexlab{b}})Li, Anand, Riess, Casertano, Yuan,
  Breuval, Macri, Scolnic, Beaton, \& Anderson}]{Li:2024pjo}
Li, S., Anand, G.~S., Riess, A.~G., {et~al.} 2024{\natexlab{b}}, Astrophys. J.,
  976, 177, \dodoi{10.3847/1538-4357/ad84f3}

\bibitem[{Li {et~al.}(2024{\natexlab{c}})Li, Jin, Li, Zhang, \&
  Zhang}]{Li:2023gtu}
Li, T.-N., Jin, S.-J., Li, H.-L., Zhang, J.-F., \& Zhang, X.
  2024{\natexlab{c}}, Astrophys. J., 963, 52, \dodoi{10.3847/1538-4357/ad1bc9}

\bibitem[{Madau \& Dickinson(2014)}]{Madau:2014bja}
Madau, P., \& Dickinson, M. 2014, Ann. Rev. Astron. Astrophys., 52, 415,
  \dodoi{10.1146/annurev-astro-081811-125615}

\bibitem[{Mastrogiovanni {et~al.}(2023)Mastrogiovanni, Laghi, Gray, Santoro,
  Ghosh, Karathanasis, Leyde, Steer, Perries, \&
  Pierra}]{Mastrogiovanni:2023emh}
Mastrogiovanni, S., Laghi, D., Gray, R., {et~al.} 2023, Phys. Rev. D, 108,
  042002, \dodoi{10.1103/PhysRevD.108.042002}

\bibitem[{Mastrogiovanni {et~al.}(2024)Mastrogiovanni, Pierra, Perri\`es,
  Laghi, Caneva~Santoro, Ghosh, Gray, Karathanasis, \&
  Leyde}]{Mastrogiovanni:2023zbw}
Mastrogiovanni, S., Pierra, G., Perri\`es, S., {et~al.} 2024, Astron.
  Astrophys., 682, A167, \dodoi{10.1051/0004-6361/202347007}

\bibitem[{Muttoni {et~al.}(2023)Muttoni, Laghi, Tamanini, Marsat, \&
  Izquierdo-Villalba}]{Muttoni:2023prw}
Muttoni, N., Laghi, D., Tamanini, N., Marsat, S., \& Izquierdo-Villalba, D.
  2023, Phys. Rev. D, 108, 043543, \dodoi{10.1103/PhysRevD.108.043543}

\bibitem[{Navarro {et~al.}(1997)Navarro, Frenk, \& White}]{Navarro:1996gj}
Navarro, J.~F., Frenk, C.~S., \& White, S. D.~M. 1997, Astrophys. J., 490, 493,
  \dodoi{10.1086/304888}

\bibitem[{Ng {et~al.}(2018)Ng, Vitale, Zimmerman, Chatziioannou, Gerosa, \&
  Haster}]{Ng:2018neg}
Ng, K. K.~Y., Vitale, S., Zimmerman, A., {et~al.} 2018, Phys. Rev. D, 98,
  083007, \dodoi{10.1103/PhysRevD.98.083007}

\bibitem[{Pascale {et~al.}(2025)}]{Pascale:2024qjr}
Pascale, M., {et~al.} 2025, Astrophys. J., 979, 13,
  \dodoi{10.3847/1538-4357/ad9928}

\bibitem[{Perivolaropoulos(2024)}]{Perivolaropoulos:2024yxv}
Perivolaropoulos, L. 2024, Phys. Rev. D, 110, 123518,
  \dodoi{10.1103/PhysRevD.110.123518}

\bibitem[{Qi {et~al.}(2022)Qi, Cui, Hu, Zhang, Cui, \& Zhang}]{Qi:2022sxm}
Qi, J.-Z., Cui, Y., Hu, W.-H., {et~al.} 2022, Phys. Rev. D, 106, 023520,
  \dodoi{10.1103/PhysRevD.106.023520}

\bibitem[{R\"as\"anen {et~al.}(2015)R\"as\"anen, Bolejko, \&
  Finoguenov}]{Rasanen:2014mca}
R\"as\"anen, S., Bolejko, K., \& Finoguenov, A. 2015, Phys. Rev. Lett., 115,
  101301, \dodoi{10.1103/PhysRevLett.115.101301}

\bibitem[{Refsdal(1964)}]{Refsdal:1964blz}
Refsdal, S. 1964, Mon. Not. Roy. Astron. Soc., 128, 307,
  \dodoi{10.1093/mnras/128.4.307}

\bibitem[{Riess {et~al.}(2019)Riess, Casertano, Yuan, Macri, \&
  Scolnic}]{Riess:2019cxk}
Riess, A.~G., Casertano, S., Yuan, W., Macri, L.~M., \& Scolnic, D. 2019,
  Astrophys. J., 876, 85, \dodoi{10.3847/1538-4357/ab1422}

\bibitem[{Riess {et~al.}(2022)}]{Riess:2021jrx}
Riess, A.~G., {et~al.} 2022, Astrophys. J. Lett., 934, L7,
  \dodoi{10.3847/2041-8213/ac5c5b}

\bibitem[{Riess {et~al.}(2024{\natexlab{a}})}]{Riess:2024vfa}
---. 2024{\natexlab{a}}, Astrophys. J., 977, 120,
  \dodoi{10.3847/1538-4357/ad8c21}

\bibitem[{Riess {et~al.}(2024{\natexlab{b}})Riess, Anand, Yuan, Casertano,
  Dolphin, Macri, Breuval, Scolnic, Perrin, \& Anderson}]{Riess:2024ohe}
Riess, A.~G., Anand, G.~S., Yuan, W., {et~al.} 2024{\natexlab{b}}, Astrophys.
  J. Lett., 962, L17, \dodoi{10.3847/2041-8213/ad1ddd}

\bibitem[{Schutz(1986)}]{Schutz:1986gp}
Schutz, B.~F. 1986, Nature, 323, 310, \dodoi{10.1038/323310a0}

\bibitem[{Sluse {et~al.}(2007)Sluse, Claeskens, Hutsemekers, \&
  Surdej}]{Sluse:2007cn}
Sluse, D., Claeskens, J.~F., Hutsemekers, D., \& Surdej, J. 2007, Astron.
  Astrophys., 468, 885, \dodoi{10.1051/0004-6361:20066821}

\bibitem[{Sluse {et~al.}(2003)Sluse, Surdej, Claeskens, Hutsemekers, Jean,
  Courbin, Nakos, Billeres, \& Khmil}]{Sluse:2003iy}
Sluse, D., Surdej, J., Claeskens, J.~F., {et~al.} 2003, Astron. Astrophys.,
  406, L43, \dodoi{10.1051/0004-6361:20030904}

\bibitem[{Song {et~al.}(2024)Song, Wang, Li, Zhao, Zhang, Zhao, \&
  Zhang}]{Song:2022siz}
Song, J.-Y., Wang, L.-F., Li, Y., {et~al.} 2024, Sci. China Phys. Mech.
  Astron., 67, 230411, \dodoi{10.1007/s11433-023-2260-2}

\bibitem[{Suyu {et~al.}(2010)Suyu, Marshall, Auger, Hilbert, Blandford,
  Koopmans, Fassnacht, \& Treu}]{Suyu:2009by}
Suyu, S.~H., Marshall, P.~J., Auger, M.~W., {et~al.} 2010, Astrophys. J., 711,
  201, \dodoi{10.1088/0004-637X/711/1/201}

\bibitem[{Suyu {et~al.}(2013)}]{Suyu:2012aa}
Suyu, S.~H., {et~al.} 2013, Astrophys. J., 766, 70,
  \dodoi{10.1088/0004-637X/766/2/70}

\bibitem[{Suyu {et~al.}(2014)}]{Suyu:2013kha}
---. 2014, Astrophys. J. Lett., 788, L35, \dodoi{10.1088/2041-8205/788/2/L35}

\bibitem[{Taylor {et~al.}(2012)Taylor, Gair, \& Mandel}]{Taylor:2011fs}
Taylor, S.~R., Gair, J.~R., \& Mandel, I. 2012, Phys. Rev. D, 85, 023535,
  \dodoi{10.1103/PhysRevD.85.023535}

\bibitem[{van Son {et~al.}(2022)van Son, de~Mink, Callister, Justham, Renzo,
  Wagg, Broekgaarden, Kummer, Pakmor, \& Mandel}]{vanSon:2021zpk}
van Son, L. A.~C., de~Mink, S.~E., Callister, T., {et~al.} 2022, Astrophys. J.,
  931, 17, \dodoi{10.3847/1538-4357/ac64a3}

\bibitem[{Wang {et~al.}(2022)Wang, Jin, Zhang, \& Zhang}]{Wang:2021srv}
Wang, L.-F., Jin, S.-J., Zhang, J.-F., \& Zhang, X. 2022, Sci. China Phys.
  Mech. Astron., 65, 210411, \dodoi{10.1007/s11433-021-1736-6}

\bibitem[{Wang {et~al.}(2018)Wang, Zhang, Zhang, \& Zhang}]{Wang:2018lun}
Wang, L.-F., Zhang, X.-N., Zhang, J.-F., \& Zhang, X. 2018, Phys. Lett. B, 782,
  87, \dodoi{10.1016/j.physletb.2018.05.027}

\bibitem[{Wang {et~al.}(2020)Wang, Zhao, Zhang, \& Zhang}]{Wang:2019tto}
Wang, L.-F., Zhao, Z.-W., Zhang, J.-F., \& Zhang, X. 2020, JCAP, 11, 012,
  \dodoi{10.1088/1475-7516/2020/11/012}

\bibitem[{Wei \& Melia(2020)}]{Wei:2020suh}
Wei, J.-J., \& Melia, F. 2020, Astrophys. J., 897, 127,
  \dodoi{10.3847/1538-4357/ab959b}

\bibitem[{Wong {et~al.}(2020)}]{H0LiCOW:2019pvv}
Wong, K.~C., {et~al.} 2020, Mon. Not. Roy. Astron. Soc., 498, 1420,
  \dodoi{10.1093/mnras/stz3094}

\bibitem[{Wu \& Zhang(2024)}]{Wu:2024faw}
Wu, P.-J., \& Zhang, X. 2024.
\newblock \doarXiv{2411.06356}

\bibitem[{Xiao {et~al.}(2024)Xiao, Shao, Wang, Song, Feng, Zhang, \&
  Zhang}]{Xiao:2024nmi}
Xiao, S.-R., Shao, Y., Wang, L.-F., {et~al.} 2024.
\newblock \doarXiv{2408.00609}

\bibitem[{Yu {et~al.}(2024)Yu, Liu, Yang, Wang, Zhang, Zhang, \&
  Zhao}]{Yu:2023ico}
Yu, J., Liu, Z., Yang, X., {et~al.} 2024, Astrophys. J. Suppl., 270, 24,
  \dodoi{10.3847/1538-4365/ad0ece}

\bibitem[{Yu {et~al.}(2020)Yu, Wang, Zhao, \& Lu}]{Yu:2020vyy}
Yu, J., Wang, Y., Zhao, W., \& Lu, Y. 2020, Mon. Not. Roy. Astron. Soc., 498,
  1786, \dodoi{10.1093/mnras/staa2465}

\bibitem[{Yun {et~al.}(2023)Yun, Han, Hu, \& Xu}]{Yun:2023ygz}
Yun, Q., Han, W.-B., Hu, Q., \& Xu, H. 2023, Mon. Not. Roy. Astron. Soc., 527,
  L60, \dodoi{10.1093/mnrasl/slad119}

\bibitem[{Zhang {et~al.}(2019{\natexlab{a}})Zhang, Zhang, Jin, Qi, \&
  Zhang}]{Zhang:2019loq}
Zhang, J.-F., Zhang, M., Jin, S.-J., Qi, J.-Z., \& Zhang, X.
  2019{\natexlab{a}}, JCAP, 09, 068, \dodoi{10.1088/1475-7516/2019/09/068}

\bibitem[{Zhang(2019)}]{Zhang:2019ylr}
Zhang, X. 2019, Sci. China Phys. Mech. Astron., 62, 110431,
  \dodoi{10.1007/s11433-019-9445-7}

\bibitem[{Zhang {et~al.}(2019{\natexlab{b}})Zhang, Wang, Zhang, \&
  Zhang}]{Zhang:2018byx}
Zhang, X.-N., Wang, L.-F., Zhang, J.-F., \& Zhang, X. 2019{\natexlab{b}}, Phys.
  Rev. D, 99, 063510, \dodoi{10.1103/PhysRevD.99.063510}

\bibitem[{Zhao {et~al.}(2011)Zhao, Van Den~Broeck, Baskaran, \&
  Li}]{Zhao:2010sz}
Zhao, W., Van Den~Broeck, C., Baskaran, D., \& Li, T. G.~F. 2011, Phys. Rev. D,
  83, 023005, \dodoi{10.1103/PhysRevD.83.023005}

\bibitem[{Zhao {et~al.}(2020)Zhao, Wang, Zhang, \& Zhang}]{Zhao:2019gyk}
Zhao, Z.-W., Wang, L.-F., Zhang, J.-F., \& Zhang, X. 2020, Sci. Bull., 65,
  1340, \dodoi{10.1016/j.scib.2020.04.032}

\bibitem[{Zheng {et~al.}(2024)Zheng, Liu, \& Qi}]{Zheng:2024mbo}
Zheng, J., Liu, X.-H., \& Qi, J.-Z. 2024, Astrophys. J., 975, 215,
  \dodoi{10.3847/1538-4357/ad7bb5}

\bibitem[{Zhu {et~al.}(2024)Zhu, Fan, Chen, Hu, \& Zhang}]{Zhu:2024qpp}
Zhu, L.-G., Fan, H.-M., Chen, X., Hu, Y.-M., \& Zhang, J.-d. 2024, Astrophys.
  J. Suppl., 273, 24, \dodoi{10.3847/1538-4365/ad5446}

\bibitem[{Zhu {et~al.}(2022{\natexlab{a}})Zhu, Hu, Wang, Zhang, Li, Hendry, \&
  Mei}]{Zhu:2021aat}
Zhu, L.-G., Hu, Y.-M., Wang, H.-T., {et~al.} 2022{\natexlab{a}}, Phys. Rev.
  Res., 4, 013247, \dodoi{10.1103/PhysRevResearch.4.013247}

\bibitem[{Zhu {et~al.}(2022{\natexlab{b}})Zhu, Xie, Hu, Liu, Li, Napolitano,
  Tang, Zhang, \& Mei}]{Zhu:2021bpp}
Zhu, L.-G., Xie, L.-H., Hu, Y.-M., {et~al.} 2022{\natexlab{b}}, Sci. China
  Phys. Mech. Astron., 65, 259811, \dodoi{10.1007/s11433-021-1859-9}

\end{thebibliography}

\end{document}